\documentclass[12pt]{article}
\usepackage{epsfig}
\usepackage{a4,isolatin1}
\usepackage{amsmath,amsfonts,latexsym, amssymb}
\usepackage{theorem}

\newtheorem{satz}{Theorem}[section]

\newtheorem{bem}[satz]{Remark}

\newtheorem{assumption}[satz]{Assumption}

\newtheorem{conclusion}[satz]{Conclusion}
\newtheorem{ob}[satz]{Observation}

\newcommand{\mbf}{\mathbf}

\newcommand{\tit}{\textit}

\newcommand{\R}{\mathbb{R}}

\begin{document}
\thispagestyle{empty}
\begin{center}
\vspace*{1.0cm}

{\LARGE{\bf Modifications of the Ornstein-Zernike \\
 Relation and the LMBW Equations \\
 in the Canonical Ensemble \\
 via Hilbert-Space Methods} \\
}

\vspace{1 cm}

{\large {\bf Manfred Requardt}}

\vspace{0.2 cm}

Institut f\"ur Theoretische Physik, Universit\"at G\"ottingen, \\
Tammannstra{\ss}e 1, 37077 G\"ottingen, Germany \\
(E-mail: requardt@theorie.physik.uni-goettingen.de)
\\[0.5cm]
{\large {\bf Heinz-J\"urgen Wagner}}

\vspace{0.2 cm}

Theoretische Physik, Universit\"at Paderborn, \\
Pohlweg 55, 33098 Paderborn, Germany \\
(E-mail: wagner@phys.uni-paderborn.de)

\end{center}

\vspace{0.5 cm}

\begin{abstract}
Application of the density functional formalism to the canonical ensemble
is of practical interest in cases where there is a marked difference
between, say, the canonical and the grand canonical ensemble (cavities
or pores). An important role is played by the necessary modification
of the famous Ornstein-Zernike relation between pair correlation and
direct correlation function, as the former is no longer invertible in
a strict sense in (finite) canonical ensembles. Here we approach the
problem from a different direction which may complement the density
functional approach. In particular, we develop rigorous canonical ensemble
versions of the LMBW equations, relating density gradient and exterior
potential in the presence of explicit (singular) containing potentials.
This is accomplished with the help of integral operator and Hilbert space
methods, yielding among other things representations of the direct
correlation function on certain subspaces. The results are particularly
noteworthy and transparent if the segregating potential is a linear
(gravitational) one. In that case the modifications in the LMBW equations
can be expressed as pure, seemingly non-local integrations over the
container boundaries.
\end{abstract} \newpage
\section{Introduction}
Recently there has been renewed interest in the the application of the
density functional method to the canonical ensemble (CE). One of the
more practical reasons is the investigation of fluids in cavities or
pores that are so small that the results depend on the choice of the
ensemble (see, {\it e.\,g.}, \cite{Evans} or, for a short review,
\cite{Loewen}). One of the methods that have turned out to be quite
useful for the description of finite systems consists of the so-called
density functional theory.

However, in its customary form this method requires the use of a
grand canonical ensemble (GCE), {\it i.\,e.}, an open system. Consequently
this restriction has led to the (re)consideration of the question in
how far the methods of density functional theory can be transferred to
the case of closed systems that have to be described in the framework
of a canonical ensemble (CE) ({\it cf.} \cite{Ramshaw} and the more recent
papers \cite{White}, \cite{White2}, \cite{White3}, \cite{Blum}).

It is the constraint of a fixed number of particles which causes certain 
technical problems. The reason is discussed in the following. We note that
these problems have already been observed in the past for several times
but, as remarked by Ramshaw \cite{Ramshaw}, have occasionally been lost 
sight of or glossed over.

The density functional method is based on the observation that for a
\tit{finite} system the external potential, $v(r)$, and the
one-particle density, $\rho(r)$, are standing in a one-to-one relation
in a gas or fluid (see \cite{Mermin}). We note however that there may
arise problems as to this property if one passes to the thermodynamic
limit and one is in a regime where the system develops long-range
correlations as, for example, in the situation of phase transitions or
spontaneous symmetry breaking.

In finite systems, however, this result leads via the functional
variation of $\rho(r)$ with respect to $v(r')$ and vice versa to an
important relation between the two-particle correlation function,
$H(r,r')$ and its inverse, $C(r,r')$, the \tit{direct
correlation function}. This relation is
frequently called the \tit{Ornstein-Zernike relation}.

It was already observed by Mermin \cite{Mermin}, that this uniqueness
relation cannot hold in the CE (in a finite volume), as the density
remains the same under a constant shift of the external potential. The
direct consequence of this fact is that
\begin{equation}\int_V H(r,r')dr'=0   \end{equation}
in the CE in a finite volume, $V$. Put differently, if we regard
$H,\,C$ as the kernels of the corresponding integral operators, ${\bf
  H}$, ${\bf C}$ on some function spaces, the constant functions are
\tit{zero eigenfunctions} of ${\bf H}$. This implies that it cannot be
straightforwardly inverted and hence the direct correlation function,
$C$, is (or rather seems to be) ill-defined (see appendix).
This is, in a nutshell, the \tit{problem of the Ornstein-Zernike
relation} in the CE.

The above phenomenon has been observed in the literature for several times,
but, nevertheless, it seems that it is by no means widely known (as has, 
for example, been also remarked in \cite{Ramshaw}).
We think it is the underlying psychological reason why people usually
prefer to use the \tit{grand canonical ensemble} without being really
aware of the above result (on the surface it is the fact that in
functional derivations one has to keep the particle number constant). 

There have been several attempts in the past to remedy this 
problem (see the papers \cite{Ramshaw}, \cite{White2}, \cite{White3},
\cite{Blum} already cited above). As far as we can see, all of them
seem to employ the functional variation method and arrive at more or
less different recipes.

We entered this field from a slightly different direction. Our main
interest centered around the long-standing and controversial issue of
the behavior of the, say, liquid-gas interface when treated from
practically first principles, that is, within the regime of
microscopic statistical mechanics. An important role in this context
is played by the famous LMBW equations and their corresponding hierarchies
\cite{LMBW}, \cite{LMBW2} (see the next sections). The underlying
common assumption in both fields is the above mentioned one-to-one
dependence between exterior field and one-particle density.

In \cite{Requwag1} we discussed in quite some detail what can happen
if there exist long-range correlations in the system or in the
interface between coexisting phases (as, for example, for vanishing
gravitational field in the liquid-gas interface). We note that one
cannot avoid these problems by making the calculations in finite
systems and passing to the thermodynamic limit afterwards.
In that case one
has to control the uniformity of convergence of various expressions in
this limit which is typically lost. For more discussion in this
direction see also \cite{Requwag2} or \cite{Requwag3} and further
references therein.

The present paper is organized as follows:  In section 2 the first LMBW
equation (the one involving the density-density correlation $H$) is
dealt with in the framework of the canonical ensemble. In particular,
the exceptional significance of boundary terms brought about by the
necessary confining part of the exterior potential is exhibited.
Moreover, for the special case where the non-confining (segregating)
part of the potential is chosen to be linear, it is shown that only
boundary integrations survive on the r.\,h.\,s of the LMBW equation
which corroborates the important role of long-ranged correlations and
the delicacy of interchanging limits.

In section 3 we derive a suitable generalization of the
Ornstein-Zernike relation for the canonical ensemble case. In contrast
to previous methods discussed in the literature, we refrain from
explicit recourses to density functional theory. Extensive use is made
instead of Hilbert space methods, the main point being the
restriction of the non-invertible integral operator ${\bf H}$ to a
subspace where it is again invertible. The correction terms for the
Ornstein-Zernike relation appearing in our approach are then compared 
with those in the previous proposals \cite{Ramshaw}, \cite{White2}, 
\cite{White3}, \cite{Blum}.

Our results allow in particular for an immediate
derivation of a (generalized) version of the second LMBW equation (the
one involving the direct correlation function $C$) for the canonical
ensemble.  This is done in section 4 and it turns out
that boundary terms have a great significance also for the case of the
second LMBW equation. In fact, whenever the non-confining part of the
exterior potential is chosen to be linear, the integrations involved
can be shown to completely reduce to boundary integrations again.

The paper is then concluded with a summary. Furthermore, an appendix 
is devoted to the fixing of some elementary notions and notations.
 
\section{The First LMBW Equation in the Canonical Ensemble}
In the following we will study systems in a finite closed container,
$V$, the walls being explicitly incorporated via appropriate
\tit{containing potentials}. We now briefly decribe how the containing
potential is fed into the respective expressions. We split the total
external potential into a \tit{segregating (bulk) potential} and a
\tit{containing potential}.
\begin{equation}v(r)=v_s(r)+v_c(r)    \end{equation}

One of the technical advantages of this method is that one can perform
the calculations as if one were in an infinite system since $v_c$
brings all quantities down to zero outside of $V$ (see
below). Furthermore, we can perform a rigorous \tit{scaling limit} of
$v_c(r)$ near the walls, thus getting an exact expression for systems
being contained by \tit{ideal hard walls}.

We start with a realistic wall potential, $v_c$, which is assumed to
be only dependent on the coordinate normal to the wall and which is
concentrated in a thin layer adjacent to the system boundary,
$\partial V$, with $v_c$ diverging smoothly to $+\infty$ at the
boundary and being $+\infty$ outside $V$. For reasons of simplicity we
illustrate the procedure with the help of a wall sitting at $x=0$, the
system being contained in the half-space $x\leq 0$.
\begin{assumption}$v_c(x)$ is assumed to be smooth for $x\leq 0$,
  $v_c(x)$ being identically zero for $x\leq -a_0<0$ and
  $v_c(x)=+\infty$ for $x\geq 0$.
\end{assumption}

An infinitely hard wall is represented by the scaling limit,
$\lambda\to\infty$, of $v_c^{\lambda}(x)=v_c(\lambda\cdot x)$. It is
instructive to control various quantities when one of the coordinates
passes through the wall. Under these circumstances the expressions are
dominated by the term $\exp{-\beta\cdot v_c(r)}$, with
$v_c(r)\to\infty$ at the wall. That is, we have for finite systems
(where all the expressions can be rigorously controlled),
\begin{equation}H(r,r')=f(r,r')\,\exp{(-\beta v_c(r'))}
\end{equation}
with $f(r,r')$ continuous across the system boundary as long as
$r$ is sitting in the bulk. The same applies to $\rho(r)$ for
$r$ passing through the wall. We have
\begin{equation}\rho(r)=f(r)\,\exp{(-\beta v_c(r))}    \end{equation}
with $f(r)$ again being continuous across the wall.

In contrast to most of the approaches to the topic described in the
introduction, we want to choose the route via a rigorous derivation of
the LMBW equations, which describe the relation between $\nabla\rho(r)$
and $\nabla v(r)$ by means of the correlation functions $H(r,r')$
and $C(r,r')$.

The LMBW hierarchies were described in \cite{LMBW}, \cite{LMBW2} with the 
help of \tit{density functional} methods. It is not made entirely explicit
what happens with boundary terms if one uses containing potentials
and/or performs the \tit{thermodynamic limit}, $N,V\to\infty$, in the
presence of possible \tit{long-range correlations}.
Concerning this particular important point, we had an interesting
discussion with R.~Lovett some time ago (\cite{Lovettpr}; see also
\cite{Lovett}). He told us that \cite{Lovett} was in part influenced
by our earlier work in \cite{Requwag1} (see in particular section 5.3
of \cite{Lovett} to get a feeling of the epistemological problems
underlying the need to perform a thermodynamic limit).

In the following we show how to cope appropriately with the possible
singular terms which may emerge in this limit. The important point is
that, in order to enclose a finite system in a container (unless one
uses for example periodic boundary conditions, which are a little bit
artificial as long as one does not perform a thermodynamic limit in
the end), one should not gloss over the necessary bounding terms,
which, as we showed rigorously, may make an effect even in this limit.

In \cite{Requwag1} we started from the so-called
Kubo-Martin-Schwinger (KMS) property of equilibrium systems
which holds in finite as well as in infinite systems. We showed that both 
the BBGKY hierarchy and the first LMBW hierarchy emerge rigorously as the 
two opposite ends of one and the same scaling procedure, as long as
certain \tit{cluster properties} are fulfilled.  If these cluster
properties are violated, extra boundary terms may survive with possible
severe consequences for concepts like the \tit{Triezenberg-Zwanzig}
formula for the surface tension.

As a particular result of \cite{Requwag1} it turns out that for finite 
systems held together by an explicit containing potential of the sort 
described above, the first LMBW equation rigorously holds in the canonical
ensemble case (provided the containing potential is correctly dealt with). 
That is, we have for the first member of the hierarchy:
\begin{equation}
\nabla\rho(r)=-\beta\int H(r,r')\,\nabla v(r')\,dr'
\end{equation}
with $v=v_s+v_c$. What interests us here is the term on the right side
including the containing potential $v_c$. It reads
\begin{equation}
-\beta\int H(r,r')\,\nabla v_c(r')\,dr'
\end{equation}

We can simulate ideal hard walls by the method described above. Using
again as illustration a wall, sitting at $x=0$ and $v_c$ being
supported in $(-a_0,0)$ with $v_c=\infty$ for $x\geq 0$, we can
rigorously perform the scaling process with
$v_c^{\lambda}(x):=v_c(\lambda x)$ and get (for the wall described
above):
\begin{equation}
-\beta\int H(r,r')\,\partial_x
  v_c^{\lambda}(r')\,dr'= -\beta\int f(r,r')\,\exp({-\beta
    v_c^{\lambda}(x)})\,\partial_x v_c^{\lambda}(x)\,dxdr'^{(d-1)}     
\end{equation}
($d$ denoting the space dimension).

Making a variable transformation, $x':=\lambda x$, we get
\begin{equation}
\int f(r,r';x'/\lambda)\partial_{x'} \exp({-\beta v_c(x')})dx'dr'^{(d-1)}
\end{equation}
where the meaning of the abbreviations should be clear. In the limit
$\lambda\to\infty$ we hence get for the $x'$-integration 
(with $v_c(-a_0)=0\;,\;v_c(0)=+\infty$):
\begin{equation}
\int f(r,r';x=0)\left(\int_{-a_0}^0\partial_x \exp{(-\beta v_c(x))}dx\right)
dr'^{(d-1)}=-\int f(r,r';x=0)\,dr'^{(d-1)}     
\end{equation}

Note that for a realistic ({\em i.\,e.}, unscaled) containing potential,
$H(r,r')\to 0$ strongly for $r'\to \partial V$ from the interior of $V$
. In the case of an ideal hard wall (scaling limit) we have in general
\begin{equation} H(r,r')\to H_+(r,r')\neq 0     \end{equation}
for $r'\to \partial V$ with the plus sign denoting the limit taken from
the interior of $V$. For $r'$ lying outside of $V$, $H(r,r')$ is of
course zero in both cases with $H_+$ usually having a jump-discontinuity.
\begin{conclusion} In case of an ideal hard wall the LMBW equation reads
\begin{equation}\nabla\rho(r)=-\beta\int_V H(r,r')\,\nabla
  v_s(r')dr'-\int_{\partial V}H_+(r,r')d\bar{o}'     \end{equation}
with $d\bar{o}'$ the outwardly oriented (vector-valued) surface element. 
Note that $H_+(r,r')$ now depends only on the segregation potential $v_s$.
\end{conclusion}
{\it Remark:} We ignore the situation near possible points of the containing
surface, $\partial V$, where the surface normal may change
discontinuously. These points do not contribute in the scaling limit
as they are of measure zero and can be avoided anyhow by making the
surface smooth.

Frequently the external gravitational potential is represented by a
linear term, $v_s(r):=mgz$, {\em viz.}
\begin{equation}\nabla v_s(r)=m\,g\,e_z    \end{equation}
($e_z$ being the unit vector in $z$-direction). While for infinite
systems one has to take into account the unboundedness from below of
such a linear term, everything is well defined for a finite volume.

Our above version of the first LMBW equation in the canonical ensemble has
the following, as we think, interesting consequence.
\begin{ob}As a result of the exact vanishing of $\int_V H(r,r')\,dr'$ and
the assumed linearity of $v_s$, we have
\begin{equation}
\nabla\rho(r)=-\beta\int H(r,r')\,\nabla v_c(r')\,dr'     
\end{equation}
or
\begin{equation}
\nabla\rho(r)=-\int_{\partial V}H_+(r,r')\,d\bar{o}'    
\end{equation}
\end{ob}
\begin{conclusion}As the containing potential is assumed to have its
  support restricted to a thin layer adjacent to the walls of the
  container, $H$ or $H_+$ must include macroscopic (long-range) density 
  correlations between the infinitesimal neighborhood of the container
  walls and the interior of the vessel as long as $\nabla\rho(r)\neq0$.
\end{conclusion}

In \cite{Lebowitz} this phenomenon was attributed to finite size
effects in the canonical ensemble. A function like $\rho^{(2)}(r,r')$
with $r'$ near or at the wall can be related to the \tit{conditional
  probability} for finding a particle at position $r$ in the bulk,
say, under the assumption that another particle is sitting at a fixed
position, $r'$, near the wall. For free systems or short range
correlations this quantity is expected to be (approximately) related
to the two-particle distribution function for a ($N$-1)-particle
system within $V$.

The consequence for the two-particle correlation function, $H(r,r')$,
is that for large separation ($r$ in the bulk, $r'$ near the wall),
{\em i.\,e.}, $|r-r'|\to\infty$ in a necessarily vague (or, rather, 
pragmatic) sense in a large but finite system, it does not go to exactly 
zero in a canonical ensemble but a finite size effect term of the order
$O(N^{-1})$ survives.

These observations therefore furnish us with an exceedingly transparent
example for which we can scrutinize some of the problems posed by Lovett
and Buff in section 5.3 of \cite{Lovett}. In particular, it can now be
clearly seen that there are indeed cases for which the structure at
remote boundaries determines the local state of a system. The main point
to be stated here is that this feature can nevertheless be brought into
accord with the option to perform a thermodynamic limit. This is due
to the extreme nonuniformity of the latter. For every single finite
system with $N$$=$\,$\mbox{const.}$ the density gradient is expressed
exclusively in terms of boundary integrations without explicit recourse to
the segregating part of the potential due to the vanishing of
$\int H(r,r')\,dr'$. On the other hand, after having performed 
the thermodynamic limit, the information previously encoded in the boundary
terms seems to have become unaccessible. However, it is by no means lost!
It has merely been shifted to the segregating potential term
which ceases to vanish in the $N$$=$\,$\infty$ case as the constant 
functions are no longer zero eigenfunctions of ${\bf H}$ there.

We already emphasized in \cite{Requwag1} the important point that the
interchange of limits in the derivation of the LMBW equation may be
a delicate matter, as the various limits might behave in a highly nonuniform
way. Here we have seen that in the canonical ensemble such nonuniform
behavior is even indispensable in order to have meaningful thermodynamic
limit states!

{\it Remark:} As was already mentioned, a linear segregating potential is
unbounded from below, and it is therefore necessary to retain parts of the
boundary walls in order to keep the system stable as $N,V$$\to$\,$\infty$.
This does however not invalidate the above considerations.

For the ideal gas these effects can be easily calculated. Without an
external potential $v_s(r)$ we have
\begin{equation}
\rho^{(2)}(r,r')-\rho(r)\rho(r')=V^{-2}(N^2-N)-V^{-2}N^2
=-(N/V)\cdot V^{-1}\sim  N^{-1}
\end{equation}
With $v_s(r)$ non-zero (and $r\neq r'$) we get:
\begin{equation}
H(r,r')=\rho^{(2)}(r,r')-\rho(r)\rho(r')
=-N\exp(-\beta(v(r)+v(r')))\cdot(\int\exp(-\beta v(r))dr)^{-2}
\end{equation}

The details of the scaling with the size of the system can be inferred
from the following set up. We take for example a rectangular
parallelepiped with area of the base equal to $A$, height equal to $h$
and with the base located at $z=0$, the exterior potential being only
$z$-dependent. We arrive at:
\begin{equation}H(r,r')=-N\cdot A^{-2}(\int\exp(-\beta
  v(z))dz)^{-2}\cdot \exp(-\beta(v(r)+v(r')))     \end{equation}
and in the particular case $v(z)=mgz$ this is equal to
\begin{equation}
-N\cdot A^{-2}\cdot (\beta\cdot mg)^2\cdot 
(1-e^{-\beta (mgh)})^{-2}\cdot\exp(-\beta(mgz+mgz'))    
\end{equation}

Inserting the corresponding terms in the expression for $H(r,r')$,
$r\neq r'$, in the \tit{grand canonical ensemble} we get on the
other hand for the ideal gas:
\begin{equation}\rho^{(2)}(r,r')-\rho(r)\rho(r')=0    \end{equation}

If we take the limit $h\to\infty$ with $N/A$ kept fixed,
we end up with the \tit{barometric formula}:
\begin{equation}H(z,0)=(N/A)\cdot A^{-1}\cdot (\beta mg)^2\exp(-\beta mgz)
\end{equation}
That is, $H$ scales as $A^{-1}$. The LMBW equation leads to
\begin{equation}
d/dz(\rho(z))=(N/A)\cdot (\beta mg)^2\cdot \exp(-\beta mgz) 
\end{equation}
or
\begin{equation}
\rho(z)=(N/A)\cdot (\beta mg)\cdot \exp(-\beta mgz)    
\end{equation}
\begin{conclusion}The local law between density and exterior
  potential is in the LMBW equation hidden in the non-local
  integration over the container walls. It can however be shown that
  in the case of the ideal gas this reduces basically to an application
  of \tit{Gauss's divergence theorem}.
\end{conclusion}
{\it Remark:} Even in the case of a zero external field and ideal hard
  walls the individual contributions of the walls on $\nabla\rho(r)$
  in the bulk are non-vanishing and of mean-field type. In a finite
  volume and the ideal gas, they however exactly compensate each other
  such that we have in fact $\nabla\rho(r)=0$ as is the case by
  calculating $\rho(r)$ directly. Once again we see from this that care has
  to be taken when performing the thermodynamic limit as the contributions
  do not decay and are independent of the distance of the respective
  walls!

\section{The Direct Correlation Function in the Can\-oni\-cal Ensemble}

The ordinary Ornstein-Zernike relation
\begin{equation}
\int C(r,r'')\,H(r'',r')\,dr'' =\delta(r-r')
\end{equation}
can be rewritten in integral operator notation as
\begin{equation}
{\bf C}\,{\bf H} = {\bf 1}
\end{equation}
where the integral operators ${\bf C}$ and ${\bf H}$
are defined as follows:
\begin{equation}
({\bf C}f)(r)=\int C(r,r')\,f(r')\,dr'\,;\quad
({\bf H}f)(r)=\int H(r,r')\,f(r')\,dr'\,.
\end{equation}
That is, ${\bf C}$ is nothing but the (left) operator inverse
of ${\bf H}$ in some properly chosen linear space. 

Some remarks are here perhaps in order concerning the reason to
introduce Hilbert spaces in the following. In contrast to, say,
quantum theory, there is no inner theoretical reason to employ the
Hilbert space formalism. The operator ${\bf H}$ lives on a certain
space of functions. As the corresponding integral kernel, $H(r,r')$
strongly decreases like $\exp(-\beta v_c(r'))$ when $r'$ approaches the
container walls, even singular functions like $\nabla v(r')$ are in
the domain of definition of ${\bf H}$ (see below). On the other hand,
typically $C(r,r')$ diverges with the same degree when $r'$ approaches
the container walls. Hence it is only defined on a smaller space of
functions than for example $L^2(V)$. We conclude that ${\bf H}$
naturally lives on a function space larger than $L^2(V)$, while 
${\bf C}$ lives on a smaller subspace of $L^2(V)$.  
In any case, ${\bf C}$ is defined on the range of ${\bf H}$ 
which is dense in $L^2(V)$ whenever ${\bf H}$ does not possess 
(proper) eigenfunctions with eigenvalue zero. However, the validity
of this latter property can only be expected if one works within the 
grand canonical ensemble.

If one works in the canonical ensemble instead and again chooses 
$\mathfrak{H}=L^2(V)$ as Hilbert space, it is immediately obvious 
that ${\bf H}$ cannot be inverted there, since---as was already pointed out 
in the introduction (see also the  appendix)---as a direct consequence of
\begin{equation}
\int H(r,r')\,dr' = 0\,
\end{equation}
it possesses the constant functions as proper eigenfunctions with 
eigenvalue zero.  

However, due to the fact that we introduced a Hilbert space of functions 
as the domain of definition for ${\bf H}$ we now have the important technical 
notions of orthogonality and symmetric operators at our disposal, and we 
hence automatically know that eigenspaces are orthogonal and left invariant by
the corresponding symmetric operator. In particular, these properties 
straightforwardly lead to the finding that a restricted form of 
invertibility for ${\bf H}$ is still present.
To demonstrate this, let us carry through the just mentioned standard
decomposition of the Hilbert space $\mathfrak{H}$ into a direct sum of
the eigenspace $\mathfrak{H}_0$ belonging to the eigenvalue zero
and its orthogonal complement~$\mathfrak{H}_0^{\bot}$:
\begin{equation}
\mathfrak{H} = \mathfrak{H}_0 \oplus \mathfrak{H}_0^{\bot}\,.
\end{equation}
As $H(r,r')=H(r',r)$, the integral operator ${\bf H}$ is symmetric in
$\mathfrak{H}$. Therefore it leaves invariant the subspace
$\mathfrak{H}_0^{\bot}$ and is invertible there. 

Under the premise that $\rho[v]$ and $\rho[\hat{v}]$
differ whenever the external potentials $v$ and $\hat{v}$ differ by
more than a constant \cite{Mermin}, it seems natural to assume
that only the constant functions are annihilated by
${\bf H}$, {\it i.\,e.}, that $\mathfrak{H}_0$ is one-dimensional.
In this case $\mathfrak{H}_0^{\bot}$ is given by
\begin{equation}
\mathfrak{H}_0^{\bot} = \left\{f\,\left|\,\int f(r)\,dr=0\right.\right\}\,.
\end{equation}
(The Hilbert spaces considered in this section consist of course of
equivalence classes of functions rather than functions itself. For
simplicity reasons we suppress this fact here.)

A problem which we already addressed is that the Hilbert space
\mbox{$\mathfrak{H}=L^2(V)$} does not exhaust the full set of
physically relevant functions on which one wants to define the
operator ${\bf H}$ ({\it cf.} for example the LMBW equations). That is,
in order to incorporate also these singular functions into the Hilbert
space formalism, which is desirable on physical grounds, we have to
modify the above ordinary Lebesgue measure.

The mentioned divergencies can, for example, be compensated by the
vanishing of the density $\rho(r)$ which roughly behaves as
$\exp(-\beta\,v_c(r))$ in the vicinity of the container boundary. That
is, one possibility consists of replacing the Hilbert space
$L^2(V)$ with the Hilbert space
$L^2(V,\rho(r)dr)$ in which the measure is now
chosen as $\rho(r)\,dr$ instead of $dr$.  While not being in
$L^2(V)$, the containing potentials and their derivatives are
now naturally contained in $L^2(V,\rho(r)dr)$.

One should however emphasize that this choice is evidently \tit{not} unique. 
Other options like, {\it e.\,g.}, the replacement of $\rho(r)$ with the 
corresponding ideal gas density 
\begin{equation}\label{id}
\rho_{\mbox{\scriptsize id}}(r)
=\frac{N\,\exp(-\beta\,v(r))}{\int\exp(-\beta\,v(r))\,dr}
\end{equation} 
or with the quantity $N(\partial\rho(r)/\partial N)_v$ 
(provided the latter expression is well-defined and positive) are possible. 
As we shall see below, this latter choice would in particular lead to results 
which are in some sense comparable with the treatment in \cite{White2}, 
\cite{White3}. 

In order to keep matters as general as possible, we therefore introduce 
the Hilbert space  
\begin{equation}
\tilde{\mathfrak{H}}=L^2(V,\tilde{\rho}(r)dr)
\end{equation}
where the measure $\tilde{\rho}(r)\,dr$ fulfils the normalization condition
\begin{equation}
\int \tilde{\rho}(r)\,dr = N\,.
\end{equation}
Moreover, we require $\tilde{\rho}(r)$ to be capable of compensating the 
singular behavior of $\nabla v$ at the container boundaries. Thus, each of 
the above-mentioned quantities $\rho,\rho_{\mbox{\scriptsize id}}, 
N(\partial\rho/\partial N)_v$ turns out to be a possible candidate 
for $\tilde{\rho}$.

As a consequence, while $H(r,r')$ is unique by its definition as a canonical 
expectation value of a certain microscopic observable and does not depend 
on the choice of the Hilbert space measure, the situation turns out to be 
quite different for $C(r,r')$. We think this observation is one of 
the virtues of our more functional-analytically oriented approach.

An additional interesting feature of the Hilbert space $\tilde{\mathfrak{H}}$
shows up if one deals with systems that still consist of a finite number of
particles $N$ but for which the volume is infinite. Such systems still
make sense, {\it e.\,g.}, in those cases where the confinement of particles
comes about by the use of polynomial wells, {\it i.\,e.}, potentials
$v(r)$ which diverge $\sim |r|^n$ as $|r|\to\infty$.

The constant functions are no longer elements of $\mathfrak{H}$ in such a
$V=\infty$ case, but the integral operator ${\bf H}$ can still be applied to
them and they are still annihilated. Therefore, the eigenvalue zero would be
an improper eigenvalue lying outside the discrete part of the spectrum.

In contrast to this, the constant functions are still contained in
$\tilde{\mathfrak{H}}$:
\begin{equation}
||1||^2 = \int \tilde{\rho}(r)\,dr = N < \infty\,.
\end{equation}
Therefore, zero is still an ordinary (proper) eigenvalue of ${\bf H}$ here.

The price one has to pay for this adaptation of the Hilbert space
consists of the fact that the integral operator ${\bf H}$ is no longer
symmetric, since
\begin{equation}
<f,{\bf H}g>\,=\int f(r)\,({\bf H}g)(r)\,\tilde{\rho}(r)\,dr
= \int f(r)\,H(r,r')\,g(r')\,\tilde{\rho}(r)\,dr'\,dr
\end{equation}
is different from
\begin{equation}
<{\bf H}f,g>\,=\int ({\bf H}f)(r')\,g(r')\,\tilde{\rho}(r')\,dr'
=\int f(r)\,H(r,r')\,g(r')\,\tilde{\rho}(r')\,dr'\,dr\,.
\end{equation}
Among other things, this leads to the unpleasant feature that the subspace
$\tilde{\mathfrak{H}}_0^{\bot}$ in the self-explanatory decomposition
\begin{equation}
\tilde{\mathfrak{H}}=\tilde{\mathfrak{H}}_0\oplus\tilde{\mathfrak{H}}_0^{\bot}
\end{equation}
is not left invariant by ${\bf H}$ which means that a straightforward
inversion is no longer possible there.

However, one immediately notices that the modified integral kernel
\begin{equation}
H'(r,r')=\frac{H(r,r')}{\tilde{\rho}(r)}
\end{equation}
leads to a symmetric integral operator ${\bf H}'$, since
\[
<f,{\bf H'}g>\,=\int f(r)\,({\bf H'}g)(r)\,\tilde{\rho}(r)\,dr
= \int f(r)\,H(r,r')\,g(r')\,dr'\,dr
\]
\begin{equation}
=\int f(r)\,H(r',r)\,g(r')\,dr'\,dr
=\int ({\bf H}'f)(r')\,g(r')\,\tilde{\rho}(r')\,dr'=\,<{\bf H}'f,g>\,.
\end{equation}
Moreover, the eigenfunctions of ${\bf H}$ and ${\bf H}'$ with eigenvalue zero
coincide. Thus, as ${\bf H}'$ is symmetric, it leaves
$\tilde{\mathfrak{H}}_0^{\bot}$ invariant. Restriction to this subspace
therefore makes an inversion of ${\bf H}'$ possible.

As the elements of $\tilde{\mathfrak{H}}_0$ are annihilated by ${\bf H}'$,
one therefore gets the following modification of the Ornstein-Zernike
relation for the canonical ensemble case:
\begin{equation}
{\bf C}'\,{\bf H}' = {\bf P}^{\bot}
\end{equation}
with ${\bf P}^{\bot}$ denoting the projection operator onto the subspace
$\tilde{\mathfrak{H}}_0^{\bot}$.
\smallskip

Since
\[
({\bf C}'\,{\bf H}'\,f)(r)=\int C'(r,r')\,H'(r',r'')f(r'')\,dr'\,dr''
\]
\begin{equation}
=\int C'(r,r')\,\frac{H(r',r'')}{\tilde{\rho}(r')}f(r'')\,dr'\,dr''\,,
\end{equation}
it is immediately clear that the direct correlation function $C(r,r')$ and
its corresponding integral operator ${\bf C}$ have to be introduced {\it via}
\begin{equation}
C(r,r')=\frac{C'(r,r')}{\tilde{\rho}(r')}\,\,.
\end{equation}
Thus, the modified Ornstein-Zernike relation for the canonical ensemble
can also be rewritten as
\begin{equation}
{\bf C}\,{\bf H} = {\bf P}^{\bot}\,.
\end{equation}

\begin{bem} \label{FS}
  While an inverse of ${\bf H}'$ does not exist outside
  $\tilde{\mathfrak{H}}_0^{\bot}$, a symmetric extension of the 
  operator ${\bf C}'$ can be defined on at least a dense subset of the full 
  Hilbert space $\tilde{\mathfrak{H}}$, namely the linear combinations of 
  elements from the range of ${\bf H}'$ (which is at least dense in 
  $\tilde{\mathfrak{H}}_0^{\bot}$) and $\tilde{\mathfrak{H}}_0$. In 
  particular, $\tilde{\mathfrak{H}}_0$ is always left invariant by such 
  an---of course non-unique---extension of ${\bf C}'$.
\end{bem}
If one again assumes that $\tilde{\mathfrak{H}}_0$ is one-dimensional,
{\it i.\,e.}, that ${\bf H}$ and ${\bf H}'$ only annihilate the constant
functions, this implies
\begin{equation}
\int C'(r,r')\,dr'= \mbox{const.}
\end{equation}
and hence we have the useful identity 
\begin{equation} \label{X}
\int C(r,r')\tilde{\rho}(r')\,dr'= \mbox{const.}
\end{equation}
Furthermore, $\tilde{\mathfrak{H}}_0^{\bot}$ is then given by
\begin{equation}
\tilde{\mathfrak{H}}_0^{\bot} =
\left\{f\,\left|\,<1,f>\,\,=\int f(r)\,\tilde{\rho}(r)\,dr=0\right.\right\}
\end{equation}
and ${\bf P}^{\bot}$ is defined as follows:
\begin{equation}
({\bf P}^{\bot}f)(r)=f(r)-\frac{<1,f>}{||1||^2}
=f(r)-\frac{1}{N}\int f(r)\,\tilde{\rho}(r)\,dr\,.
\end{equation}
The modification of the Ornstein-Zernike relation for the canonical ensemble
can therefore be expressed as
\begin{equation}
\int C(r,r'')\,H(r'',r''')f(r''')\,dr''\,dr'''
=f(r)-\frac{1}{N}\int f(r)\,\tilde{\rho}(r)\,dr\,.
\end{equation}
If one replaces $f(r)$ with the delta function $\delta(r-r')$
concentrated at an arbitrary but fixed point $r'$, the following final
form emerges:

\begin{conclusion}
A modified Ornstein-Zernike relation in the canonical ensemble can be
written down as follows:
\begin{equation} \label{Y}
\int C(r,r'')\,H(r'',r')\,dr''=\delta(r-r')-\frac{\tilde{\rho}(r')}{N}\,.
\end{equation}
\end{conclusion}

Roughly speaking, our version of a modified Ornstein-Zernike relation
differs from the one given in \cite{White2}, \cite{White3} by the
appearance of the term $\tilde{\rho}(r')/N$ instead of 
$(\partial\rho(r')/\partial N)_v$. 

To be more specific, let us compare our relations (\ref{X}), (\ref{Y}), 
{\it i.\,e.},
\begin{equation} 
\int C(r,r'')\,H(r'',r')\,dr''+\frac{\tilde{\rho}(r')}{N}=\delta(r-r')\,,
\end{equation}
\begin{equation} \label{YZ} 
\int C(r,r')\,\frac{\tilde{\rho}(r')}{N}\,dr'= \mbox{const.}
\end{equation}
with the generalized Ornstein-Zernike relations
that have been given as equations (52) and (53) in \cite{White3}:
\begin{equation} \label{ZZ}
\int C(r,r'')\,H(r'',r')\,dr''
+\left(\frac{\partial\rho(r')}{\partial N}\right)_{\!\!v}
=\delta(r-r')\,,
\end{equation}
\begin{equation} \label{Z}
\int C(r,r')\left(\frac{\partial\rho(r')}{\partial N}\right)_{\!\!v}dr'= 
\beta\left(\frac{\partial\mu}{\partial N}\right)_{\!\!v}.
\end{equation}
The difference of these latter relations to ours above is twofold. 
On the one hand it consists in the already mentioned replacement
of $\tilde{\rho}(r')/N $ with $(\partial\rho(r')/\partial N)_v$. This
difference reflects the freedom of choice for the measure of the
Hilbert space $\tilde{\mathfrak{H}}$ in our approach and would become
inexistent with the particular definition $\tilde{\rho}(r)\,dr$ $=$
$N(\partial\rho(r)/\partial N)_v\,dr$ (provided the expression on 
the r.\,h.\,s.~is a well-defined positive measure at all). On the 
other hand it turns out that the constant on the r.\,h.\,s.~of (\ref{Z}) 
is no longer arbitrary but has a definite value (corresponding to a definite
extension of the operator ${\bf C}'$). Both differences can therefore
be characterized as an implication of some freedom of choice in our
approach which is not present in that of refs.~\cite{White2} and
\cite{White3}. 

A brief remark is here in order concerning the results in \cite{Ramshaw}, 
\cite{Blum} and their relation to that of \cite{White2}, \cite{White3}. 
We think that in both approaches the guiding idea is to mollify the canonical 
ensemble in order to make it more akin to the grand canonical case. 
This can be achieved in, on the surface, slightly different but nevertheless 
closely related ways. The strategy pursued in \cite{White2}, \cite{White3}
is to embed the canonical ensemble with fixed particle number $N$ into an 
enlarged parameter space, thus making a variation with respect to $N$ 
possible (see, however, the related short discussion in our summary 
at the end of the paper). Another possibility is to make the system 
resemble an open system by truncating the weakly decaying tails (coming 
from the constancy of $N$) in the correlation functions as in \cite{Ramshaw} 
or \cite{Blum}. Both strategies lead to closely interconnected results 
({\it cf.}~section 3.4 of~\cite{White3}) and correspond, as we 
have seen, to a particular choice in our own approach.

However, nonwithstanding the fact that equations (\ref{ZZ}), (\ref{Z})
are the natural outcome of the inversion formalism of the extended
variable approach in \cite{White2} and \cite{White3}, it should
be stressed that there is no obligation to make the
corresponding choices for the measure and the constant also in our
approach. Quite to the contrary, there is a certain amount of freedom
which allows us to adopt some considerably simpler expressions.
Furthermore, in the following we shall frequently deal with
expressions where the operator ${\bf C}'$ is solely applied to
elements from the range of ${\bf H}'$. In these cases we need not even
worry about definite extensions of ${\bf C}'$.  It is therefore
obvious that---contrary to $H(r,r')$---the direct correlation function
$C(r,r')$ is by no means a uniquely defined quantity in our approach.
This has the advantage that it is to some extent possible to simplify
matters by using ``tailor-made'' definitions for $C$.

To further clarify the contents of the present chapter, it might be
useful to deal with the ideal gas as a relatively simple but 
thereby also fully tractable example. The expressions 
$\rho(r)\,dr$, $\rho_{\mbox{\scriptsize id}}(r)\,dr$, and 
$N(\partial\rho(r)/\partial N)_v\,dr$ all coincide in this case 
and represent an obvious choice for our measure $\tilde{\rho}\,dr$. 
In the canonical ensemble we have
\begin{equation}
H_{\mbox{\scriptsize id}}(r,r')
=\rho_{\mbox{\scriptsize id}}(r)\,\delta(r-r')
-\frac{1}{N}\,\rho_{\mbox{\scriptsize id}}(r)\,
\rho_{\mbox{\scriptsize id}}(r')\,.
\end{equation}
With $\tilde{\rho}=\rho_{\mbox{\scriptsize id}}$ we are therefore led to
\begin{equation}
H'_{\mbox{\scriptsize id}}(r,r')
=\frac{H_{\mbox{\scriptsize id}}(r,r')}{\rho_{\mbox{\scriptsize id}}(r)}
=\delta(r-r')-\frac{\rho_{\mbox{\scriptsize id}}(r')}{N}
\end{equation}
which means that $\tilde{\mathfrak{H}}_0$ is indeed one-dimensional
and that ${\bf H}'_{\mbox{\scriptsize id}}$ is simply given as
\begin{equation}
{\bf H}'_{\mbox{\scriptsize id}}={\bf P}^{\bot}.
\end{equation}
After restriction to the subspace $\tilde{\mathfrak{H}}_0^{\bot}$,
we thus get ${\bf H}'_{\mbox{\scriptsize id}}={\bf 1}$ and its inverse
is trivially given as ${\bf C}'_{\mbox{\scriptsize id}}={\bf 1}$ there.
This immediately leads to 
\begin{equation}
C'_{\mbox{\scriptsize id}}(r,r')=\delta(r-r')\,.
\end{equation}
For the time being, this expression is only valid on 
$\tilde{\mathfrak{H}}_0^{\bot}$. However, its validity can 
immediately be extended to the full Hilbert space 
$\tilde{\mathfrak{H}}$ (see remark \ref{FS}). 
Thus, in this case the direct correlation function reads
\begin{equation}
C_{\mbox{\scriptsize id}}(r,r')
=\frac{C'_{\mbox{\scriptsize id}}(r,r')}{\rho_{\mbox{\scriptsize id}}(r')}
=\frac{\delta(r-r')}{\rho_{\mbox{\scriptsize id}}(r)}\,\,.
\end{equation}
Hence we ended up with exactly the same expression as for the grand 
canonical ensemble. Of course one has again to keep in mind here 
that the above extension of ${\bf C}'_{\mbox{\scriptsize id}}$ from
$\tilde{\mathfrak{H}}_0^{\bot}$ to $\tilde{\mathfrak{H}}$ is by no means 
a unique one. So it should not come as a surprise that, {\it e.\,g.}, 
the direct correlation function $C_{\mbox{\scriptsize id}}(r,r')$ arrived at 
in equation (73) of \cite{White3} differs from our expression by a constant.

As a further result we notice that the range of 
${\bf H}'_{\mbox{\scriptsize id}}$ consists of the full subspace 
$\tilde{\mathfrak{H}}_0^{\bot}$. This in turn allows for a definition
of ${\bf C}'_{\mbox{\scriptsize id}}$ on the full Hilbert space 
$\tilde{\mathfrak{H}}$. Thus, the reservation ``on at least a 
dense subset'' made in remark \ref{FS} is unnecessary here.

It is therefore interesting to deal with the question as to how far 
this desirable property is also transferable to the nonideal case. It
will turn out in the following that a positive answer can be given for a 
rather large class of systems if we make the special choice 
$\tilde{\rho}(r)=\rho(r)$. For a thorough discussion of the mathematics 
involved the reader may consult, {\it e.\,g.}, ref.~\cite{Reed}.

For $\tilde{\rho}(r)=\rho(r)$ we have (see appendix):
\begin{equation}\label{H}
H'(r,r')=\frac{H(r,r')}{\rho(r)}=\delta(r-r')+h(r,r')\,\rho(r')\,.
\end{equation}
This corresponds to the natural decomposition 
\begin{equation} \label{N}
{\bf H}'={\bf 1}+{\bf h}'
\end{equation}
with
\begin{equation}
({\bf h}'f)(r)=\int h(r,r')\,f(r')\,\rho(r')\,dr'\,.
\end{equation}
We now require ${\bf h}'$ to be a Hilbert-Schmidt operator on 
$\tilde{\mathfrak{H}}$, {\it i.\,e.}, 
\begin{equation}
\int|h(r,r')|^2\,\rho(r)\,\rho(r')\,dr\,dr'<\infty\,.
\end{equation}
Under this proviso the Hilbert-Schmidt theorem tells us that 
${\bf h}'$ is a completely continuous symmetric operator possessing a
purely discrete spectrum with zero as its only possible limit
(accumulation) point. Thus, the symmetric operator 
${\bf H}'={\bf 1}+{\bf h}'$ also has a purely discrete spectrum, but now 
with 1 as its only possible limit point. As an immediate consequence, 
the inverse of ${\bf H}'$ is defined on the full subspace
$\tilde{\mathfrak{H}}_0^{\bot}$ because the critical eigenvalue of
${\bf H}'$ is the eigenvalue zero which, according to our preceding
discussion, can at most be finitely degenerated. The above result is 
also a particular example for the so-called ``Fredholm alternative''
which states that whenever ${\bf k}$ is a completely continuous
operator on a Hilbert space, then either $({\bf 1}+{\bf k})^{-1}$ is
defined everywhere or ${\bf 1}+{\bf k}$ possesses an eigenvector
with eigenvalue zero. 

It should finally be remarked that the simple form of the above natural 
decomposition (\ref{N}) of ${\bf H}'$ highly depends on the particular 
choice of $\varrho(r)\,dr$ as our Hilbert space measure. For any other 
choice of the measure a nontrivial multiplication operator would show up 
instead of the ${\bf 1}$, thus making an analogous reasoning as above more 
complicated. In particular, equation (\ref{H}) had to be replaced with 
\begin{equation}
H'(r,r')=\frac{H(r,r')}{\tilde{\rho}(r)}
=\sqrt{\frac{\rho(r)}{\tilde{\rho}(r)}}
\left(\delta(r-r')+\sqrt{\frac{\rho(r)\rho(r')}
{\tilde{\rho}(r)\tilde{\rho}(r')}}\,h(r,r')\,\tilde{\rho}(r')\right)
\sqrt{\frac{\rho(r')}{\tilde{\rho}(r')}} 
\end{equation}
and it turns out that the further discussion would crucially depend on the 
properties of the multiplication operator $\sqrt{\rho/\tilde{\rho}}\,$ in 
the general case.

Thus, again we see that matters can be kept simple due to the 
freedom of making the most suitable choice for the Hilbert space measure.

\section{The Second LMBW Equation in the Canonical Ensemble}

Now we are ready to examine what expression replaces the second LMBW
equation (the one including the direct correlation function) when the
canonical ensemble is used instead of the grand canonical one. 

As was already pointed out in section 2, the first LMBW equation
\begin{equation}
({\bf H}\,\nabla v)(r)=\int H(r,r')\,\nabla v(r') dr'
=-\frac{1}{\beta}\,\nabla\rho(r)
\end{equation}
is still valid when the canonical ensemble is used. Due to the
containing part of the potential, the derivatives of $v(r)$ are
strongly divergent in the vicinity of the container boundary. The
replacement of $\mathfrak{H}$ with $\tilde{\mathfrak{H}}$ therefore 
finds its justification here. Again putting
$\tilde{\mathfrak{H}}=L^2(V,\tilde{\rho}(r)dr)$, we get
\begin{equation}
({\bf H}'\,\nabla v)(r)
=-\frac{1}{\beta}\,\frac{\nabla\rho(r)}{\tilde{\rho}(r)}\,.
\end{equation}

The operator ${\bf H}'$ transforms
elements of $\tilde{\mathfrak{H}}$ into elements of
$\tilde{\mathfrak{H}}_0^{\bot}$, and this is reflected in
the fact that
\begin{equation}
<1,\frac{\nabla\rho}{\tilde{\rho}}>\,=\int\nabla\rho(r)\,dr=0
\end{equation}
due to the vanishing of $\rho(r)$ at the container boundary.

The operator ${\bf C}'$ can safely be applied onto both sides of the above
LMBW equation and this leads to
\begin{equation}
({\bf C}'\,{\bf H}'\,\nabla v)(r)
=-\frac{1}{\beta}\,\Big({\bf C}'\,\frac{\nabla\rho}{\tilde{\rho}}\Big)(r)
=-\frac{1}{\beta}\,({\bf C}\,\nabla\rho)(r)\,.
\end{equation}
Employing the modified Ornstein-Zernike relation
${\bf C}'\,{\bf H}'={\bf P}^{\bot}$, we are thus immediately led to
\begin{equation}
({\bf C}\,\nabla\rho)(r)= -\beta\,({\bf P}^{\bot}\nabla v)(r)
\end{equation}
or, if $\tilde{\mathfrak{H}}_0$ is again assumed to be one-dimensional, to
the following

\begin{conclusion}
The second LMBW equation in the canonical ensemble can be written as
\begin{equation}
\int C(r,r')\,\nabla\rho(r')\,dr'=-\beta\left(\nabla v(r)
-\frac{1}{N}\int\tilde{\rho}(r)\,\nabla v(r)\,dr\right)\,.
\end{equation}
\end{conclusion}

Therefore---in contrast to the unaltered first LMBW equation---this
second one may differ from its grand canonical counterpart. The
difference, however, ceases to exist for those choices of the measure
where ${\bf P}^{\bot}\nabla v = \nabla v$, {\em i.\,e.}, where the
components of $\nabla v$ are already elements
of $\tilde{\mathfrak{H}}_0^{\bot}$. 

An example for this latter property is provided by the choice 
$\tilde{\rho}(r)=\rho_{\mbox{\scriptsize id}}(r)$ 
as equation (\ref{id}) immediately implies
\begin{equation}
\int\rho_{\mbox{\scriptsize id}}(r)\,\nabla v(r)\,dr
=-\frac{1}{\beta}\int\nabla\rho_{\mbox{\scriptsize id}}(r)\,dr=0
\end{equation}
due to the vanishing of $\rho_{\mbox{\scriptsize id}}$ at the
container boundary.

There is, however, an even more remarkable example:
  
\begin{ob} Within the canonical ensemble we have
\begin{equation}
\int\rho(r)\,\nabla v(r)\,dr=0\,.
\end{equation}
Hence, for the choice $\tilde{\rho}(r)=\rho(r)$ the components of 
$\nabla v(r)$ already lie in the orthogonal subspace 
$\tilde{\mathfrak{H}}_0^{\bot}$. Since $\nabla v$ contains both the 
segregating and the containing potential, we also have
\begin{equation}
\int\rho(r)\,\nabla v_s(r)\,dr=-\int\rho(r)\,\nabla v_c(r)\,dr.
\end{equation}
\end{ob}\smallskip 

{\it Proof:} The explicit inclusion of the containing potential greatly
facilitates the calculations as we can perform all manipulations in
the full $\R^n$ (with the Boltzmann weight being zero for all
configurations with the position of at least one particle lying
outside the volume $V$). We assume that the interaction potential
$U(r_1,\ldots,r_N)$ between the $N$ particles is translation
invariant. By $v_a$ we denote the exterior potential translated by an
arbitrary vector $a \in \R^n$: 
\begin{equation}
v_a(r)=v(r+a)\,.
\end{equation}
Under these assumptions a simple variable transform shows that the canonical 
partition function $Q_N[v]$ remains invariant under translations of $v$: 
\begin{equation}
Q_N[v]=Q_N[v_a]\,.
\end{equation}
The same holds for the logarithm and hence for the free energy.
Differentiation with respect to $a$ at $a=0$ immediately yields
the desired result. \hfill {\it q.\,e.\,d.} 

\medskip

The above result can also be shown in a slightly different way
by employing the following equation which---again for the case of a
translation invariant interaction potential---states that the time derivative 
of the momentum density must have zero expectation value in thermal
equilibrium ({\em cf.}, for example, \S\,2 of \cite{Schofield}): 
\begin{equation}
\mbox{Div}\;\!\mbox{\boldmath$\pi$}(r) + \rho(r)\,\nabla v(r) = 0\,.
\end{equation}
Due to the appearance of the divergence of the pressure tensor 
$\mbox{\boldmath$\pi$}(r)$ in this equation, $\int\rho(r)\,\nabla v(r)\,dr$ 
can be reexpressed as a surface integral. Shifting the surface of 
integration beyond the container boundaries one again sees that 
this integral is in fact zero.

The physical ramifications of this result are interesting. They show
that the condition of equilibrium entails an a priori compensation of
the global force exerted by the segregating potential on the system
and the forces of constraint, provided by the container walls and
acting on the nearby elements of the fluid or gas. This can most
easily be seen in the case of the ideal gas (see our above
calculations).

Moreover, it is immediately seen that in those cases where well-behaved 
differentiation of $\int\rho(r)\nabla v(r)\,dr$ with respect to $N$ is 
possible, we also have
\begin{equation}
\int\left(\frac{\partial\rho(r)}{\partial N}\right)_{\!\!v}
\nabla v(r)\,dr=0\,.
\end{equation}
Thus, whenever $\tilde{\rho}(r)=N(\partial\rho(r)/\partial N)_v$
is a possible choice, we have a further case for which
${\bf P}^{\bot}\nabla v=\nabla v$. 

In concluding this section we just give one little example for the 
implications which are brought about by the above results.
Let the external potential as usual be divided into a segregating and a
containing part. In particular, we choose the segregating part to be
a linear (gravitational) potential :
\begin{equation}
v(r)=v_s(r)+v_c(r)=m\,g\!\cdot\!r+v_c(r)\,.
\end{equation}
(Here and in the following $g$ denotes a constant vector which can,
{\it e.\,g.}, be used to define the $z$-direction.)

Then, due to
\begin{equation}
\nabla v_s(r) -\frac{1}{N}\int\tilde{\rho}(r)\,\nabla v_s(r)\,dr
=mg-\frac{mg}{N}\,\int\tilde{\rho}(r)\,dr=mg-mg=0\,,
\end{equation}
the segregating part of $v(r)$ drops out from the right hand side
of the second LMBW equation and we are left with
\begin{equation}
({\bf C}\,\nabla\rho)(r)=-\beta\left(\nabla v_c(r)
-\frac{1}{N}\int\tilde{\rho}(r)\,\nabla v_c(r)\,dr\right)\,.
\end{equation}
(Strictly speaking, the segregating potential remains of course still present
to some extent in those cases where we choose a $\tilde\rho$ that 
implicitly or explicitly depends on it.)

The containing potential $v_c(r)$ can be chosen such that it
identically vanishes for $r$ outside an arbitrarily small vicinity of the
container boundary ${\partial V}$. If $r$ is sufficiently distant from
this boundary, it therefore lies outside the support of $v_c$ and the last
equation reduces to
\begin{equation}
({\bf C}\,\nabla\rho)(r)=
\frac{\beta}{N}\int\tilde{\rho}(r)\,\nabla v_c(r)\,dr\,.
\end{equation}

As in the grand canonical ensemble, where
$({\bf C}\,\nabla\rho)(r) = -\beta m g$ for $r$ sufficiently distant from
the boundary, $({\bf C}\,\nabla\rho)(r)$ is still constant in the
canonical case, but the value of this constant can be quite different in 
general. With special choices such as, {\em e.\,g.},
\begin{equation}
\tilde{\rho}(r)=\frac{N\,\exp(-\beta\,v_c(r))}{\int\exp(-\beta\,v_c(r))\,dr}
\end{equation}
it can even be made to vanish!

Moreover, as we have already seen,
\begin{equation}
\frac{1}{N}\int\tilde{\rho}_0(r)\,\nabla v(r)\,dr
= m g + \frac{1}{N}\int\tilde{\rho}_0(r)\,\nabla v_c(r)\,dr=0
\end{equation}
whenever $\tilde{\rho}_0$ denotes any of the expressions
$\rho,\rho_{\mbox{\scriptsize id}},N(\partial\rho/\partial N)_v$.
This allows us to express the difference with the
grand canonical case somewhat more explicitly:
\begin{equation}
({\bf C}\,\nabla\rho)(r)=-\beta m g +\frac{\beta}{N}\int\Big(\tilde{\rho}(r)
-\tilde{\rho}_0(r)\Big)\,\nabla v_c(r)\,dr\,.
\end{equation}

This leads to the interesting fact that the whole information about the
value of $({\bf C}\,\nabla\rho)(r)$ in the canonical ensemble as well as
about its possible difference with the corresponding grand canonical ensemble 
value $-\beta m g$ is contained in the values of certain functions on the 
support of $\nabla v_c$, {\it i.\,e.}, in an extremely small vicinity of the
container boundary.

This last statement can even be made somewhat more explicit by the
observation that one of course can also make the transition from
smooth containing potentials to ideal hard walls here. Proceeding in
an analogous way as in section~2, we are left with
\begin{equation}
({\bf C}\,\nabla\rho)(r)=
\frac{1}{N}\int_{\partial V} \tilde{\rho}_+(r)\,d\bar{o}
=-\beta m g+\frac{1}{N}\int_{\partial V}
(\tilde{\rho}_+(r)-\tilde{\rho}_{0\,+}(r))\,d\bar{o}
\end{equation}
where the index ``$+$'' once more denotes the limit taken from the
interior of $V$.

\section{Summary}

In the preceding sections we developed a method to establish a kind of
Ornstein-Zernike relation in the canonical ensemble which is complementary 
to the more common density functional approach. In the canonical ensemble
the main problem is that the two-particle correlation function, when viewed 
as an integral operator $\mbf{H}$, inevitably has a zero eigenvalue and thus 
cannot be straightforwardly inverted. Its inverse $\mbf{C}$, representing 
the direct correlation function, therefore seems to be ill-defined.

Several suggestions have been made in the past to deal with or to circumvent
this problem. In contrast to most of the treatments, we emphasize
an explicit operator theoretic point of view. We show that $\mbf{H}$
can be inverted on some appropriate subspace of a certain function
space being endowed with an adapted scalar product and being
orthogonal to the constant functions which represent the eigenfunctions
with eigenvalue zero. We introduce several intermediary function spaces
by means of which it becomes particularly transparent that (and moreover 
in what precise sense) $\mbf{C}$ is non-unique on the full function space, 
the non-uniqueness stemming both from the chosen particular intermediary 
function space and the genuine non-uniqueness of the operator extension 
from a subspace to the full function space.

We note that these intermediary spaces are practically unavoidable if one 
wants to apply $\mbf{H}$ and $\mbf{C}$ to their natural space of functions 
which comprise also singular functions like, {\it e.\,g.}, $\nabla v_c(r)$, 
$v_c$ denoting the containing potential confining the system to a finite 
container. These latter functions occur in the LMBW equations which represent 
a natural context for such relations and which we therefore treated as a 
straightforward application. To put it in a nutshell, while $H$ is genuinely 
unique due to its explicit expression in statistical mechanics, this is not
so for $C$. Different intermediary spaces and different extensions
lead to slightly different expressions for the direct correlation
function which however are all fulfilling a restricted Ornstein-Zernike 
relation for the canonical ensemble. 

A connection of our approach with the ones developed earlier in the realm 
of density functional theory can be made in those cases where the quantity
$(\partial\rho(r)/\partial N)_v$ is well-defined and positive such that
$N(\partial\rho(r)/\partial N)_v\,dr$ is a possible choice for the measure 
of our Hilbert space. 

In this context it should perhaps be pointed out that the uniqueness
or even well-definedness of quantities involving partial derivatives 
with respect to $N$ is by no means self-evident. While the formal 
differentiation with respect to $N$ can be given a qualitative meaning in 
the realm of phenomenological thermodynamics ({\it i.\,e.}, in some kind of 
thermodynamic limit where one can make use of the equivalence of ensembles), 
we think the situation is much less clear whenever one wants to generate 
such expressions without transcending the realm of the canonical ensemble. 
This is even more so if one aims at the description of systems 
where the particle number is so small that the difference of ensembles 
becomes perceptible. The main problem occurring here is that the particle 
number $N$ shows up in the dimension of the phase space volume and 
thus in the phase space measure as basically a discrete number. It is 
therefore not entirely evident how one should technically define unique
continuous derivatives with respect to $N$ without dismissing the 
confinement to the canonical ensemble case. This problem was also 
observed by Ray \cite{Ray} who even claims that there is no useful 
way to carry out such differentiations in general.

One of the virtues of our approach therefore consists in the possibility 
to circumvent the above problem by choosing simpler measures such as 
$\rho(r)\,dr$ or $\rho_{\mbox{\scriptsize id}}(r)\,dr$ for our Hilbert space.
Furthermore, differentiation with respect to $N$ also plays a decisive role 
for the constant on the r.\,h.\,s.~of equation (\ref{Z}). However, the 
non-uniqueness of the direct correlation function in our approach also
enables us to alter this constant. It is therefore again possible 
to avoid differentiation with respect to $N$ here.

As a further advantageous feature of the mentioned non-uniquenesses 
we finally note that the corresponding freedom of choice quite often 
allows for certain simplifications of procedures.

\section*{Appendix}
For the convenience of the reader and in order to fix the notation we
provide a short list of general relations. 
The \tit{pair correlation functions}, $H(r,r'),\;h(r,r')$, are
defined as
\begin{equation}
H(r,r'):= \rho^{(2)}(r,r')+ \rho(r)\,\delta(r-r')- \rho(r)\,\rho(r')\,, 
\end{equation}
\begin{equation}
h(r,r'):=\frac{\rho^{(2)}(r,r')-\rho(r)\,\rho(r')}{\rho(r)\,\rho(r')}
=:g(r,r')-1  
\end{equation}
with the ordinary molecular distribution functions 
$\rho(r)$, $\rho^{(2)}(r,r')$ (see for example \cite{Hill}).
The physical meaning of $H(r,r')$ can be inferred from the relation
\begin{equation}
H(r,r')=\langle\delta n(r)\,\delta n(r')\rangle
\end{equation}
with $\langle\ldots\rangle$ denoting the ensemble expectation value and 
$\delta n(r)$ being the true microscopic observable
\begin{equation}
\delta n(r):=n(r)-\langle n(r)\rangle=n(r)-\rho(r)\,,\quad
n(r):=\sum_i\delta(r-r_i)\,.
\end{equation}
This identity exhibits the fluctuation content of $H(r,r')$.

The \tit{Ornstein-Zernike relation} which in its general form can
only hold in the grand canonical ensemble (see below) defines the 
\tit{direct correlation function} and reads
\begin{equation} \label{OZ}
\int C(r,r')\,H(r',r'')\,dr'=\delta(r-r'') 
\end{equation}
or with
\begin{equation}
C(r,r'):=\delta(r-r')/\rho(r)-c(r,r')
\end{equation}
we have
\begin{equation}
h(r,r'')-c(r,r'')=\int h(r,r')\rho(r')c(r',r'')dr'
=\int c(r,r')\rho(r')h(r',r'')dr'\,.    
\end{equation}

Working in the canonical ensemble in a finite volume $V$ and inserting the
expressions for $\rho^{(2)}$ and $\rho$, we get the remarkable identity
\begin{equation}
\int_V H(r,r')\,dr'=(N-1)\,\rho(r)+\rho(r)-N\rho(r)=0\,.    
\end{equation}
In other words, the integral operator ${\bf H}$ with kernel $H(r,r')$ 
annihilates the constant functions and therefore cannot be inverted. 
The Ornstein-Zernike relation in its usual form (\ref{OZ}) is thus 
unavailable in the canonical ensemble case.

\section*{Acknowledgment} 

We thank the two referees for a series of valuable remarks and suggestions.



\begin{thebibliography}{99}
{\small
\bibitem{Evans} R.~Evans: ``Fluids in Model Pores or Cavities: The Influence
of Confinement on Structure and Phase Behaviour'', NATO Science Series C 529
(1999) 153
\bibitem{Loewen} H.~Loewen: ``Density Functional Theory of Inhomogeneous
Classical Fluids'', J.~Phys.~Cond.~Matter 14 (2002) 11897
\bibitem{Ramshaw} J.~D.~Ramshaw: ``Functional Derivative Relations for a
Finite Non-Uniform Molecular Fluid in the Canonical Ensemble'',
Mol.~Phys.~41 (1980) 219
\bibitem{White} J.~A.~White, A.~Gonz\'alez, F.~L.~Rom\'an, S.~Velasco,
``Density-Functional Theory of Inhomogeneous Fluids in the Canonical
Ensemble'', Phys.~Rev.~Lett.~84 (2000) 1220
\bibitem{White2} J.~A.~White, S.~Velasco: ``The Ornstein-Zernike Equation
in the Canonical Ensemble'', Europhys.~Lett.~54 (2001) 475
\bibitem{White3} J.~A.~White, A.~Gonz\'alez: ``The Extended Variable Space
Approach to Density Functional Theory in the Canonical Ensemble'',
J.~Phys.~Cond.~Matter~14 (2002) 11907
\bibitem{Blum} J.~A.~Hernando, L.~Blum: ``Density Functional Formalism in
the Canonical Ensemble'', J.~Phys.~Cond.~Matter 13 (2001) L577
\bibitem{Mermin} N.~D.~Mermin: ``Thermal Properties of the Inhomogeneous
Electron Gas'', Phys.~Rev.~A 137 (1965) 1441
\bibitem{LMBW} R.~Lovett, C.~Y.~Mou, F.~P.~Buff: ``The Structure of the
Liquid-Vapor Interface'', J.~Chem.~Phys.~65 (1976) 570
\bibitem{LMBW2} M.~S.~Wertheim: ``Correlations in the Liquid-Vapor Interface'',
J.~Chem.~Phys.~65 (1976) 2377
\bibitem{Requwag1} M.~Requardt, H.~J.~Wagner: ``(Infinite) Boundary
Corrections for the LMBW-Equations and the TZ-Formula of Surface
Tension in the Presence of Spontaneous Symmetry Breaking'',
Physica A 154 (1988) 183
\bibitem{Requwag2} M.~Requardt, H.~J.~Wagner: ``Does the Three-Dimensional
Capillary Wave Model Lead to a Universally Valid and Pathology-Free
Description of the Liquid-Vapor Interface near $g=0$? A Controversial Point
of View'', J.~Stat.~Phys.~64~(1991)~807
\bibitem{Requwag3} M.~Requardt, H.~J.~Wagner: ``Implications of a
Non-Analytic Small-$k$ Asymptotics of the Direct Correlation Function
in the Liquid-Vapor Interface'', Physica A 196 (1993) 59
\bibitem{Lovettpr} R.~Lovett: private communication
\bibitem{Lovett} R.~Lovett, F.~P.~Buff: ``Examples of the Construction of
Integral Equations in Equilibrium Statistical Mechanics from Invariance
Principles'', Physica A 172 (1991) 147
\bibitem{Lebowitz} J.~L.~Lebowitz: ``Asymptotic Value of the Pair
Distribution Near a Wall'', Phys.~Fluids 3 (1960) 64
\bibitem{Reed} M.~Reed, B.~Simon: {\em Methods of Modern Mathematical 
Physics I}\,, Academic Press, New York - San Francisco - London, 1972
\bibitem{Schofield} P.~Schofield, J.~R.~Henderson: ``Statistical
Mechanics of Inhomogeneous Fluids'', Proc.~Roy.~Soc.~London A 379 
(1982) 231
\bibitem{Ray} J.~R.~Ray: ``Correct Boltzmann Counting'', 
Eur.~J.~Phys.~5 (1984) 219
\bibitem{Hill} T.~L.~Hill: {\em Statistical Mechanics}\,, 
Dover Publications, New York, 1987
}
\end{thebibliography}
\end{document}